\documentclass[aps,twocolumn,showpacs,floatfix,superscriptaddress]{revtex4}
\usepackage[dvips]{graphicx}
\usepackage{dcolumn}
\usepackage{amsmath}

\begin{document}

\title{High Precision Measurement of the Thermal Exponent for the three-dimensional XY Universality Class}
\author{Evgeni Burovski}
\author{Jonathan Machta}
\affiliation{Department of Physics, University of Massachusetts,
Amherst, MA 01003}
\author{Nikolay Prokof'ev}
\author{Boris Svistunov}
\affiliation{Department of Physics, University of Massachusetts,
Amherst, MA 01003}%
\affiliation{Russian Research Center ``Kurchatov Institute'',
123182 Moscow, Russia}

\begin{abstract}
Simulations results are reported for critical point of the
two-component $\phi^4$ field theory.  The correlation length
exponent is measured to high precision with the result
$\nu=0.6717(3)$. This value is in agreement with recent simulation
results [Campostrini \textit{et al}., Phys. Rev. B \textbf{63},
214503 (2001)], and marginally agrees with the most recent
space-based measurements of the superfluid transition in $^4$He
[Lipa \textit{et al}., Phys. Rev. B \textbf{68}, 174518 (2003)].
%
\end{abstract}


\maketitle Universality at critical points is among the most
beautiful and powerful ideas to emerge from statistical
physics~\cite{Fisher}.  The unversality hypothesis asserts that
critical exponents and some other asymptotic properties of
critical points are independent of microscopic details and depend
on only a few properties such as the symmetry of the order
parameter and spatial dimensionality. Universality permits the
calculation of critical exponents for experimental systems using
simplified and optimized model systems with the same symmetries.
The theory of critical phenomena also predicts relationships
amongst critical exponents.  Among these relations are the
hyperscaling relation, $\alpha = 2-d\nu$,  between the specific
heat exponent $\alpha$, the correlation length exponent $\nu$ and
the dimensionality $d$ and the Josephson relation, $\zeta =
(d-2)\nu$ for the superfluid stiffness exponent $\zeta$.
Improvements in experiments, theory and computer simulations have
led to increasingly strict tests of universality and scaling
relations.    By far the most accurate experimental measurements
of critical exponents are for the superfluid transition in $^4$He,
which is in the $O(2)$ or XY universality class.  The purity of
liquid Helium together with the stability of temperature control
and the accuracy of specific heat measurements at low temperatures
means that the limiting factor in approaching the critical
singularities is the rounding due to the Earth's gravitational
field.
To overcome gravitational rounding, space-based microgravity
experiments have been devised \cite{Lipa03} that achieve four
significant digit accuracy for the specific heat exponent
$\alpha$.  These experimental results seemingly did not agree with
either analytical or numerical calculations. The experimental
value itself evolved with time due to reanalysis of the data, see
Refs.\ \cite{Lipa96,Lipa00,Lipa03}. The most recent reported value
being $\nu_\mathrm{exp}=0.6709(1)$, as inferred from the measured
value of $\alpha$ via the hyperscaling relation.

A number of recent analytical and numerical calculations were
aimed at high precision determinations of $\nu$ for the XY  
universality class. The vortex-loop calculations by
Williams~\cite{Williams} yield $\nu=0.6717$. T\"or\"ok and
Hasenbusch~\cite{HasTokok1999} studied the two-component $\phi^4$
model via Monte Carlo simulations.  This model has a parameter
that adjusts the softness of the potential constraining amplitude
fluctuations of the order parameter.  These authors took advantage
of this freedom to suppress the leading corrections to scaling.
This study resulted in $\nu=0.6723(3)(8)$ (the statistical and
systematic errors are given in the first and second bracket,
respectively). Later the effort was advanced by Campostrini
\textit{et al}. \cite{Camp01} who combined Monte Carlo simulations
with a high-temperature expansion to obtain $\nu=0.67155(27)$,
which agrees with the experimental result at the level of two
standard deviations. Our study also employs a $\phi^4$ model with
fine-tuning of the Hamiltonian.

\begin{figure}[htb]
\includegraphics[width=\columnwidth,keepaspectratio=true]{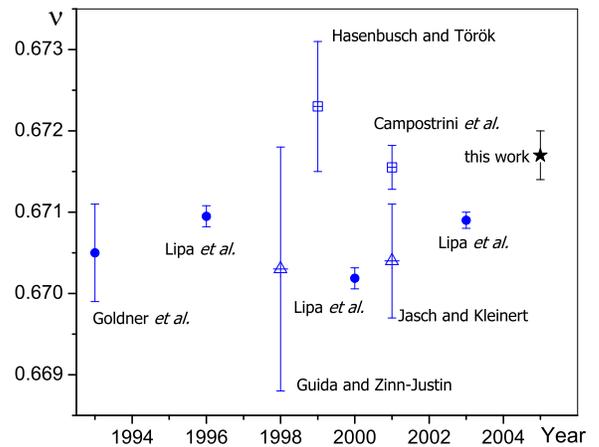}
\caption{Results for $\nu$ as a function of time. Filled circles
show the experimental values, open triangles depict the
field-theoretic calculations, and squares show the Monte Carlo
results.}
 \label{fig-nu}
\end{figure}

Purely analytic studies of the XY 
critical point include extensive treatments by Guida and
Zinn-Justin, summarized in Ref. \cite{Guida98}, which yield
$\nu=0.6703(15)$ for the perturbative seven-loop expansion in
$d=3$, and $\nu=0.6680(35)$ for the $\epsilon$-expansion up to the
order $\epsilon^5$. Jasch and Kleinert \cite{JaschKleinert2001}
developed a Borel resummation technique in the context of a
strong-coupling theory which results in $\nu=0.6704(7)$.  The
history of recent results for the correlation length exponent is
given in Fig.~\ref{fig-nu}.

The purpose of this paper is to provide high precision simulations
results for the exponent $\nu$ for the 3D XY 
universality class using only the Monte Carlo simulations and to
compare with other theoretical, computational and experimental
results.

We study a discrete 3D classical $\phi^4$ model, defined by the
Hamiltonian
\begin{equation}
\frac{H}{T} = -t\sum_{\langle ij \rangle} \phi_{i}^{*} \phi_{j} + %
     (U/2) \sum_{i} |\phi_{i}|^4 - \mu \sum_{i} | \phi_{i} |^2, %
\label{ham}
\end{equation}
where $i$ and $j$ label sites of a simple three-dimensional cubic
lattice with periodic boundary conditions,  $\langle ij \rangle$
stands for the pairs of nearest neighbor sites, and $\phi_i$ is
the complex order parameter field. Since $\phi_i$ fields are
continuous and unbounded, the model (\ref{ham}) has only two
independent parameters: by rescaling fields one can always set one
parameter to unity.  Henceforth, we set $\mu=1$.


In order to simulate the model (\ref{ham}), we employ the
high-temperature expansion for the partition function which
transforms the configuration space into that of closed oriented
loops. The latter can be efficiently sampled by the worm algorithm
\cite{worm}, which switches between the partition function and
Green function sectors. The worm algorithm has virtually no
critical slowing down and provides a direct access to the
statistics of winding number fluctuations, which, in turn, define
the superfluid stiffness~\cite{Ceperley87}:
%
%
\begin{equation}
\rho_s L =\sum_{a=1}^d \langle W_a^2\rangle /2d.
\label{rho_s}
\end{equation}

Here $L$ is the linear system size, $W_a$ is the winding
number in the $a$-th direction, and angular brackets denote
averaging over the Gibbs distribution.
One can also devise direct Monte Carlo estimators for the
derivatives of $\rho_s$ with respect to $t$ and/or $U$, which involve
cross-correlations between energy and winding numbers.

In the renormalization group (RG) framework, the finite-size
scaling of the superfluid stiffness obeys the relation
\begin{equation}
\rho_s L = f(x) + g_{\omega}(x) y_{\omega}L^{-\omega}+\dots,
 \label{RG1}
\end{equation}
where $x=(L/\xi)^{1/\nu}$ is the dimensionless scaling variable,
$\xi=\xi(t,U)$ is the correlation length, $f(x)$ and $g(x)$ are
universal functions that are analytic as $x \to 0$,
$y_{\omega}$ is the leading irrelevant scaling field, and
dots represent the higher-order corrections.
Field-theoretical calculations \cite{Guida98} yield $\omega =
0.802(11)$ ($\epsilon$-expansion) and $\omega=0.789(11)$ ($d=3$
loop expansion), and the numerical analysis of Ref.\ \cite{Camp01}
gives $\omega=0.795(9)$.

By differentiating Eq.\ (\ref{RG1}) with respect to $t$ for
$U=U_c$ and then letting $t \to t_c$, one transforms (\ref{RG1})
into
\begin{equation} R'_c =  A L^{1/\nu}(1 + C L^{-\omega}) +
\text{higher-order terms}.%
\label{RG2}
\end{equation}
Here the derivative of $R \equiv \rho_s L$ is taken \emph{at} the
critical point, and $A$ and $C$ are non-universal constants. Eq.\
(\ref{RG2}) is especially convenient for the numerical data
analysis: the log-log plot of $R'_c$ versus $L$ is a straight line
with the slope $1/\nu$ for sufficiently large $L$. The second term
adds a slight concavity or convexity to the curve for intermediate
system sizes, depending on the sign of $C$.
 It is argued in Refs. [6,7] that there is little advantage
in using improved models if corrections to scaling are
 included in the fits of MC data, and two alternative
 strategies of dealing with the problem are suggested. One is
 that MC data are fitted by discarding correction-to-scaling
 terms, and the possible systematic error thus introduced
 is estimated from the universal amplitude ratios. The other
 is that MC data for the critical point are used in the
 analysis of the high-temperature expansion series. We demonstrate
 below that comparable accuracy can be achieved by Monte Carlo
 alone, using joint fits of several improved models.



In order to locate a particular critical point, we fix some value
of $t$ and then plot $R = \rho_s L$ as a function of $U$ for
different system sizes. The crossing of these curves in the limit
of $L \to \infty$ gives the critical point \cite{Binder}.

We first consider the critical point $t_\mathcal{A} = -0.0795548$
and $U_\mathcal{A} = 0.4101562(14)$, which we denote by
$\mathcal{A}$.
%
%
This critical point was previously studied by Campostrini
\textit{et al.} \cite{Camp01} who performed an extensive and
thorough search of improved models \footnote{the notation of Ref.
\cite{Camp01} is related to our's via $t=\beta/2(2\lambda-1)$ and
$U=2\lambda/(2\lambda-1)$.}.

The accuracy of the critical point determination is verified by an
independent Monte Carlo measurement of the first and second
derivatives of $R$. Indeed, suppose that for a given $U=U_c$ the
value of $t$ is off by $\delta t = t - t_c$. Then, expanding Eq.\
(\ref{RG1}) up to $O(\delta t^2)$ and using the data from Table
\ref{table1} one makes sure that (i) the deviations of $R_c$ from
its universal value $0.2580(3)$ are consistent with $\delta t \sim
10^{-6}$, and (ii) the $O(\delta t^2)$ terms are smaller than
statistical errors and, thus, can be safely neglected. For
example, $R_\mathcal{A}(L=48)=0.2583(1)$ and
$R''_\mathcal{A}(L=48)=2.65\times 10^{4}$. The derivatives for $L
\neq 48$ can be computed using $R' \propto L^{1/\nu}$ and $R''
\propto L^{2/\nu}$, cf. Eq.\ (\ref{RG1}).


Table \ref{table1}  lists the raw data for $\rho_s' L$ (see below
for the discussion of the dataset $\mathcal{B}$). Each data point
was obtained from not less than $5 \times 10^8$ sweeps (upon
equilibrating) and the accumulated data set represents
approximately $18$ years of CPU time.  Error bars are one standard
deviation and were obtained using the blocking method.

\begin{table}[!tb]
\caption{Results for critical points $\mathcal{A}$ and $\mathcal{B}$.\label{table1}}%
\begin{ruledtabular}
\begin{tabular}{lll}
\multicolumn{3}{c}{
\begin{tabular}{ldd}
\\
$L$ & R'_\mathcal{A} & R'_\mathcal{B}\\
 \hline \\
4 & 2.0329(9)  & 1.9906(3) \\
5 & 2.8414(5)  & 2.7842(7) \\
6 & 3.7316(4)  & 3.6583(9) \\
7 & 4.6955(5)  & 4.6060(5) \\
8 & 5.7289(4)  & 5.6215(9) \\
9 & 6.8265(7)  & \text{---} \\
10 & 7.9848(9) & 7.839(1) \\
11 & 9.2031(13) & \text{---} \\
12 & 10.474(2) & 10.284(1) \\
16 & 16.074(4) & 15.784(2) \\
20 & 22.403(4) & 22.011(3) \\
24 & 29.396(7) & 28.883(3) \\
32 & 45.095(13) & 44.33(1) \\
48 & 82.48(3)& 81.05(2) \\
64 & \text{---} & 124.40(4) \\
96 & 231.56(17) & \text{---} \\
\end{tabular}
}\\ 
%
\end{tabular}
\end{ruledtabular}
\end{table}

\begin{figure}[htb]
\includegraphics[width=0.99\columnwidth,keepaspectratio=true]{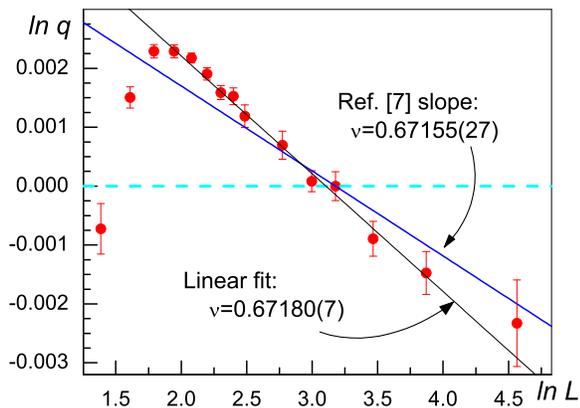}
\caption{The derivative of $R_\mathcal{A}$ rescaled via Eq.\
(\ref{RG_rescale}). Dashed horizontal line corresponds to
$\nu=\nu_\mathrm{exp}$. See the text for discussion.}
 \label{fig-lin}
\end{figure}

In order to magnify the fine details, we scale the numerical
data with the experimental exponent,
\begin{equation}
q(L) = \frac{R'_c(L)L^{-1/\nu_{\mathrm{exp}}}}{R'_c(L_0)
L_0^{-1/\nu_{\mathrm{exp}}}}
\label{RG_rescale}
\end{equation}
and normalize it at an arbitrarily chosen value, $L_0=24$. Figure
\ref{fig-lin} shows the data for critical point $\mathcal{A}$,
rescaled using Eq.\ (\ref{RG_rescale}). Given the error bars, it
appears that corrections-to-scaling are relevant only for $L
\lesssim 10$ and for $L \gtrsim 10$ a straight line with the slope
equal to $1/\nu -1/\nu_\mathrm{exp}$ is a good fit, see Eqs.\
(\ref{RG2},\ref{RG_rescale}). A linear fit of the data for $L \geq
10$ yields $\nu_\mathcal{A} = 0.67180(7)$, in a flagrant
disagreement with the experimental value
$\nu_\mathrm{exp}=0.6709(1)$. Our value $\nu_\mathcal{A}$ is in
agreement with the value $\nu=0.67155(27)$ of Ref.\ \cite{Camp01}
within the combined error bars. However, our new data for critical
point $\mathcal{A}$ are more accurate and yield a smaller error
bar. We have also studied a version of the improved link-current
model \cite{j-curr}, which belongs to the same universality class.
When these data were fit by a straight line, we observed even
greater discrepancies with $\nu_{\mathrm{exp}}$.

%
%
To understand the apparent disagreement with experiment, we
performed a series of simulations for critical points in the
vicinity of the critical point $\mathcal{A}$ to map out the region
where $C=C(t_c,U_c)$ is small. We then carried out a large-scale
simulation of one of the critical points, $\mathcal{B}$,
characterized by $t_\mathcal{B}=-0.07142822$ and
$U_\mathcal{B}=0.3605750(8)$.
Table \ref{table1} shows the raw data for this critical point
($R_\mathcal{B}(L=48)=0.2577(1)$ and
$R''_\mathcal{B}(L=48)=1.27\times 10^{4}$). A na\"{\i}ve linear
fit of the dataset $\mathcal{B}$ for $L \geq 10$ yields
$\nu_\mathcal{B}=0.67142(5)$, which is inconsistent with
$\nu_\mathcal{A}$. Critical points $\mathcal{A}$ and $\mathcal{B}$
are equally legitimate and we are left in a quandary as to whether
to reject universality or to consider a more careful analysis of
the data.  We choose the later course.

It becomes obvious that there is no easy way to improve the
accuracy of the critical exponent calculations by ignoring
corrections to scaling, even when they appear very small. To
reconcile the results for the datasets $\mathcal{A}$ and
$\mathcal{B}$ one \emph{must} include the subleading corrections
in the fits. If this is done for each dataset separately, it leads
to a large increase (almost by an order of magnitude) in the
uncertainty of the fits, as has been noted in Ref. \cite{Camp01}.
Tighter error bars can be obtained performing  a \emph{joint} fit
of two datasets. More specifically, we perform a $6$-parameter fit
according to
\begin{equation}
\ln q \sim B + \left( \frac{1}{\nu} - \frac{1}{\nu_\mathrm{exp}}%
\right) \ln( L / L_0 ) + C L^{-\omega} ,
\label{fit_eq}
\end{equation}
as follows from Eqs.\ (\ref{RG2}) and (\ref{RG_rescale}) for $C \ll
1$. The fitting parameter $B$ is introduced in order to undo the
artificial normalization $q(L=L_0)=1$. Here each critical point
has its own amplitudes $B$ and $C$, while the universal
exponents $\nu$ and $\omega$ are shared between the datasets,
which yields six fitting parameters.

\begin{figure}[tb]
\includegraphics[width=0.99\columnwidth,keepaspectratio=true]{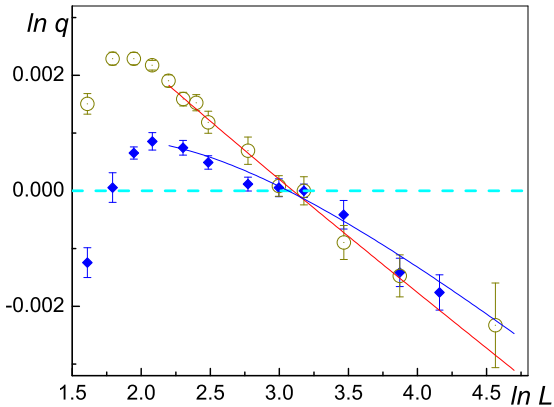}
\caption{Data from Table \ref{table1} rescaled via Eq.\
(\ref{RG_rescale}). Circles: dataset $\mathcal{A}$, diamonds:
dataset $\mathcal{B}$. Dashed horizontal line corresponds to
$\nu=\nu_\mathrm{exp}$. Solid lines are the results of the joint
fit of both datasets (see the text for explanation). Note the
scale of the vertical axis.}
 \label{fig-fits}
\end{figure}

We performed the joint fit according to Eq.\ (\ref{fit_eq}) via
stochastic minimization of $\chi$-square. We constrained the
exponent $\omega$ to within $\pm 0.03$  around $0.795$, which is
the $3\sigma$ interval of the established value \cite{Camp01}. We
also discarded small system sizes $L<L_\mathrm{cutoff}=10$. The
optimization procedure yields $\nu = 0.6717(3)$ at the confidence
level \cite{nr} $\mathrm{CL} \approx 0.43$. This value is
consistent with the result of Ref.\ \cite{Camp01}, and also
marginally agrees with the experimental value \cite{Lipa03}. The
fit is depicted in Fig.\ \ref{fig-fits} and yields
\begin{equation}
\left\{%
\begin{aligned}
C_\mathcal{A} &= (1.5 \pm 0.5 )\times 10^{-3}, \\
C_\mathcal{B} &= (-1.0 \pm 0.3)\times 10^{-2}, \\
\nu &= 0.6717(3). \\
\end{aligned}%
\right. %
\label{final}
\end{equation}
Note that the dataset $\mathcal{A}$ points are exactly the same as
in Fig.\ \ref{fig-lin} and a slight curvature is visible in Fig.
\ref{fig-fits} which is unaccounted for in Fig. \ref{fig-lin}.
It is also worth noting that the best-fit value of $\omega$ is
$\omega=0.81(1)$ which is consistent with the field-theoretic
estimates \cite{Guida98}. The error bars in Eq.\ \ref{final}
reflect both stochastic and systematic uncertainties, the latter
being estimated via changing the $L_\mathrm{cutoff}$ for the
dataset $\mathcal{A}$ and/or $\mathcal{B}$.

One might question the relevance of higher order corrections to
scaling omitted from Eq.\ (\ref{RG1}). The next order correction
is proportional to the square of the leading irrelevant field,
$\propto C^2 L^{-2\omega}$, i.e. it contains the square of the
already small amplitude $C$ \cite{Hasenb99}. A sizable cumulative
amplitude of higher-order corrections would exhibit itself for the
smallest system sizes, $L \lesssim 12$. In the contrary, the
scaling curves for both models under consideration have overall
vertical scale on the order of $10^{-3}$ for system sizes as small
as $L=4$.

The apparent controversy concerning the 3D XY 
 universality class is resolved. Our calculated value of
$\nu=0.6717(3)$ is in agreement with the results of recent
simulation \cite{Camp01}, and also marginally agrees with the best
experimental value $\nu_\mathrm{exp}=0.6709(1)$, deduced from
$\alpha$ via hyperscaling relation. We demonstrate that one has to
be careful when working with improved models and always include
corrections to scaling into the fit, even when the data set allows
a good linear fit. The fourth-digit accuracy in critical exponents
can be reached by simultaneously fitting more than one critical
point.


We acknowledge support from NASA under Grant NAG-32870.


\end{document}